\documentclass[]{elsart}

\usepackage{graphicx,amsmath,amssymb}

\begin{document}

\begin{frontmatter}

\title{Growing community networks with local events}

\author{Xin-Jian Xu$^{a,b}$},
\author{Xun Zhang$^{c}$},
\author{J. F. F. Mendes$^{b}$}

\address{$^{a}$Department of Mathematics, College of Science, Shanghai University, Shanghai 200444, China}
\address{$^{b}$Departamento de F\'{i}sica da Universidade de Aveiro, 3810-193 Aveiro, Portugal}
\address{$^{c}$Centre for Computational Science and Engineering, National University of
Singapore, Singapore 117542}

\begin{abstract}
The study of community networks has attracted considerable attention
recently. In this paper, we propose an evolving community network
model based on local processes, the addition of new nodes
intra-community and new links intra- or inter-community. Employing
growth and preferential attachment mechanisms, we generate networks
with a generalized power-law distribution of nodes' degrees.
\end{abstract}

\begin{keyword}
Complex Networks; Community Networks
\end{keyword}

\end{frontmatter}


\section{Introduction}

Complex networks, evolved from the Erd\"{o}s-R\'{e}nyi random graph
\cite{Erdos_59}, are powerful models for describing many complex
systems in biology, sociology, and technology \cite{Albert_02}. In
the past decade, the explosion of the general interest in the
structure and the evolution of most real-world networks is mainly
reflected in two striking characteristics. One is the small-world
property \cite{Watts_98}, which suggests that a network has a highly
degree of clustering like regular networks and a small average
distance among any two nodes similar to random networks. The
small-world phenomenon has been successfully described by network
models with some degree of randomness \cite{Watts_98,Newman_99}. The
other is the scale-free behavior \cite{Barabasi_99}, which means a
power-law distribution of connectivity, $P(k)\sim k^{- \gamma}$,
where $P(k)$ is the probability that a node in the network has $k$
connections to other nodes and $\gamma$ is a positive real number
determined by the given network. The origin of the scale-free
behavior has been traced back to two mechanisms that are observed in
many systems, growing and preferential attachment
\cite{Barabasi_99,Krapivsky_00}.

Recently, with the progress of research in networks, many other
statistical characteristics of networks appeared on the stage. Of
particular renown is the so-called \lq\lq community\rq\rq (or \lq\lq
modularity\rq\rq). That is to say, a network is composed of many
clusters of nodes, where the nodes in the same cluster are highly
connected, while there are few links among the nodes belonging to
different clusters. For instance, groups are formed in scientific
collaboration networks \cite{Girvan_02}. Also, it has been found
that dynamical processes on networks are affected by community
structures, such as tendencies spread well within communities
\cite{Bettencourt_03} and diffusion between different communities is
slow \cite{Eriksen_03}.

In the study of community networks, most research has been directed
in two distinct directions. On the one hand, attention has been paid
to designing algorithms for detecting community structures in real
networks. A pioneering method was made by Girvan and Newman
\cite{Girvan_02}, who introduced a quantitative measure for the
quality of a partition of a network into communities. Later, a
number of algorithms have been proposed in order to find a good
optimization with the least computational cost. The fastest
available procedures use greedy techniques \cite{Newman_04} and
extremal optimization \cite{Duch_05}, which are capable of detecting
communities in large networks. On the other hand, research has
focused on modeling of networks with community structures. In Ref.
\cite{Watts_02}, a static social network was introduced where
individuals belong to groups that in turn belong to groups of groups
and so on. In Ref. \cite{Gronlund_04}, a networked seceder model was
suggested to illustrate group formation in social networks. In Ref.
\cite{Noh_05}, a growing bipartite network for social communities
with group structures was proposed. Each of those models is
constructed based on one aspect of reality.

In this paper, we introduce a network model with communities that
gives a realistic description of local events
\cite{Albert_00,Dorogovtsev_00,Li_03}. The model incorporates three
processes, the addition of new nodes intra-community and new links
intra- or inter-community. Using growing and preferential attachment
mechanisms, we generate the community network with a good
right-skewed distribution of nodes' degrees, which has been observed
in many social systems.

\section{Model}

The Barab\'{a}si-Albert network \cite{Barabasi_99} only describes a
particular type of evolving networks, the addition of new nodes
preferential connecting to the nodes already present in the network.
Systems in the real world, however, are much richer. For example, in
scientific collaboration networks, a multidisciplinary scientist is
not only collaborate with scientists in his research fields but also
has a stronger desire to collaborate with scientists in other
fields. In friendship networks, a person usually makes friends with
people belonging to different communities besides the community he
belongs to. To give a realistic description of the network
construction like that, we introduce a growing model of community
networks based on local events, the addition of new nodes
intra-community and new links intra- or inter-community. The
proposed model is defined as follows.

We start with $M$ ($\ge 2$) isolated communities and each community
consists of a small number $n$ of isolated nodes. At each time step,
we perform one of the following three operations.

(i) With probability $p$ we add a new node in a randomly chosen
community. Here the randomly chosen means that the community is
selected according to the uniform distribution. The new node is only
connected to one node that already present in the selected
community. We denote it as the $u$th commnuity. The probability that
node $i$ in community $u$ will be selected is proportional to its
intra-community degree
\begin{equation}
\prod(k_{i}^{\text{intra}})=
\frac{k_{u,i}^{\text{intra}}+1}{\sum_{j}(k_{u,j}^{\text{intra}}+1)},
\label{nodeprob}
\end{equation}
where the sum runs over nodes in community $u$ and
$k_{u,i}^{\text{intra}}$ is the intra-community degree of node $i$
in community $u$.

(ii) With probability $q$ we add a new link in a randomly chosen
community. For this we randomly select a node in a randomly chosen
community $u$ as the starting point of the new link. The other end
of the link is selected in the same community with the probability
given by Eq. (\ref{nodeprob}).

(iii) With probability $r$ $(=1-p-q)$ we add a new link between two
communities. For this we randomly select a node in a randomly chosen
community $u$ as the starting point of the new link. The other end
$i$ of the link selected in the other community $v$ is proportional
to its inter-community degree
\begin{equation}
\prod(k_{i}^{\text{inter}})= \frac{k_{v,i}^{\text{inter}}+1}{\sum_{v
\neq u;j}(k_{v,j}^{\text{inter}}+1)}, \label{linkprob}
\end{equation}
where the sum runs over nodes in all communities except for
community $u$ and $k_{v,i}^{\text{inter}}$ is the inter-community
degree of node $i$ in community $v$.

After $t$ time steps, this scheme generates a network of $Mn+pt$
nodes and $t$ links. The parameters $p$, $q$, and $r$ control the
network structure. In the case of small $r$, the generated network
will have a strong community structure. Notice that whatever process
is chosen in the network growth, only one link is added to the
system at each time step (duplicate and self-connected edges are
forbidden), however, this is not essential. We choose link
probabilities $\prod(k^{\text{intra}}_{i})$ and
$\prod(k^{\text{inter}}_{i})$ to be proportional to
$k^{\text{intra}}_{i}+1$ and $k^{\text{inter}}_{i}+1$, respectively,
such that there is a nonzero probability of isolated nodes acquiring
new links.

\section{Degree distribution}

In our community network, the degree of a node consists of two
parts, the intra-community degree and the inter-community degree.
Increase in the node's connectivity can be divided into two
processes, the increases of the intra-community degree and the
inter-community degree. In each process, we assume that
$k_{i}^{\text{intra}}$ and $k_{i}^{\text{inter}}$ change
continuously, and the probabilities $\prod(k_{i}^{\text{intra}})$
and $\prod(k_{i}^{\text{inter}})$ can be interpreted as the rates at
which $k_{i}^{\text{intra}}$ and $k_{i}^{\text{inter}}$ change,
respectively. Thus, the operations (i)-(iii) all contribute to
$k_{i}$, each being incorporated in the continuum theory as follows.

(i) Addition of a new node in a randomly chosen community with
probability $p$ :
\begin{equation}
\frac{\partial k_{u,i}^{\text{intra}}}{\partial t}=
p\frac{1}{M}\frac{k_{u,i}^{\text{intra}}+1}{\sum_{j}(k_{u,j}^{\text{intra}}+1)}.
\end{equation}

(ii) Addition of a new link in a randomly chosen community with
probability $q$ :
\begin{equation}
\frac{\partial k_{u,i}^{\text{intra}}}{\partial t}=
q[\frac{1}{N}+\frac{1}{M}\frac{k_{u,i}^{\text{intra}}+1}{\sum_{j}(k_{u,j}^{\text{intra}}+1)}],
\end{equation}
where $N$ is the number of total nodes. The first term on the
right-hand side (rhs) corresponds to the random selection of one end
of the new link, while the second term on the rhs reflects the
preferential attachment (Eq. (\ref{nodeprob})) used to select the
other end of the link.

(iii) Addition of a new links between two communities with
probability $r$ :
\begin{equation}
\frac{\partial k_{v,i}^{\text{inter}}}{\partial t}=
r[\frac{1}{N}+(1-\frac{1}{M})\frac{k_{v,i}^{\text{inter}}+1}{\sum_{v
\neq u;j}(k_{v,j}^{\text{inter}}+1)}].
\end{equation}
The first term on the rhs represents the random selection of one end
of the new link, while the second term on the rhs considers the
preferential attachment (Eq. (\ref{linkprob})) used to select the
other end of the link in the other community.

Combing the contribution of above processes, we have
\begin{eqnarray}
\frac{\partial k_{u,i}^{\text{intra}}}{\partial t} &=&
\frac{p+q}{M}\frac{k_{u,i}^{\text{intra}}+1}{\sum_{j}(k_{u,j}^{\text{intra}}+1)}+\frac{q}{N}, \label{evolve1}\\
\frac{\partial k_{v,i}^{\text{inter}}}{\partial t} &=&
\frac{r}{N}+r\frac{M-1}{M}\frac{k_{v,i}^{\text{inter}}+1}{\sum_{v
\neq u;j}(k_{v,j}^{\text{inter}}+1)}, \label{evolve2}
\end{eqnarray}
with
\begin{eqnarray}
\sum_{j}(k_{u,j}^{\text{intra}}+1) &=&
\sum_{j}k_{u,j}^{\text{intra}}+\frac{N}{M} \notag\\
&=& 2t(p\frac{1}{M}+q\frac{1}{M})+\frac{Mn+pt}{M} \notag\\
&=& \frac{3p+2q}{M}t+n, \notag\\
\sum_{v \neq u;j}(k_{v,j}^{\text{inter}}+1) &=&
\sum_{v \neq u;j}k_{v,j}^{\text{inter}}+N(1-\frac{1}{M}) \notag\\
&=& 2tr\frac{M-1}{M}+(Mn+pt)\frac{M-1}{M} \notag\\
&=& \frac{(2-p-2q)(M-1)}{M}t+(M-1)n. \notag
\end{eqnarray}
We can simplify Eqs. (\ref{evolve1}) and (\ref{evolve2}) for large
$t$
\begin{eqnarray}
\frac{\partial k_{u,i}^{\text{intra}}}{\partial t} &\approx&
\frac{p+q}{3p+2q}\frac{(k_{u,i}^{\text{intra}}+1)}{t}+\frac{q}{pt}, \label{eqintra}\\
\frac{\partial k_{v,i}^{\text{inter}}}{\partial t} &\approx&
\frac{1-p-q}{2-p-2q}\frac{(k_{v,i}^{\text{inter}}+1)}{t}+\frac{1-p-q}{pt}.
\label{eqinter}
\end{eqnarray}
The boundary conditions of the intra-community degree and the
inter-community degree at initial time $t_s$ can be estimated in the
sense of mathematical expectations,
$k_{u,i}^{\text{intra}}(t_{s})=p+q$ and
$k_{v,i}^{\text{inter}}(t_{s})=r$, respectively. So we write the
solutions of Eqs. (\ref{eqintra}) and (\ref{eqinter})
\begin{eqnarray}
k_{u,i}^{\text{intra}}(t) & = &
\frac{p^3+p^2+2p^2q+4pq+pq^2+2q^2}{p(p+q)}(\frac{t}{t_{s}})^{\frac{p+q}{3p+2q}} - \frac{p^2+4pq+2q^2}{p(p+q)}, \label{formulaintra}\\
k_{v,i}^{\text{inter}}(t) & = &
\frac{2-2q-pq+p-p^2}{p}(\frac{t}{t_{s}})^{\frac{1-p-q}{2-p-2q}} -
\frac{2-2q}{p}. \label{formulainter}
\end{eqnarray}
In random networks, the degree distribution can be calculated by
\begin{equation}
P(k)=\frac{1}{t}\sum_{i=1}^{t}\delta(k_i(t)-k),
\end{equation}
which gives
\begin{eqnarray}
P(k^{\text{intra}}) & = &
\frac{3p^2+2pq}{p^2+2q^2+4pq+2p^2q+pq^2+p^3} \notag\\
&\times&
\left[\frac{p^2+4pq+2q^2+(p^2+pq)k^{\text{intra}}}{p^2+2q^2+4pq+2p^2q+pq^2+p^3}\right]^{-(3+\frac{p}{p+q})},
\label{degreeintra}\\
P(k^{\text{inter}}) & = &
\frac{2p-2pq-p^2}{2-p-4q-2p^2+2q^2+2p^2q+pq^2+p^3} \notag\\
&\times&
\left[\frac{2-2q+pk^{\text{inter}}}{2+p-2q-pq-p^2}\right]^{-(3+\frac{p}{1-p-q})}.
\label{degreeinter}
\end{eqnarray}
Thus, the degree distribution of our network obeys a generalized
power-law form
\begin{equation}
P(k) \sim [A(p,q)k + B(p,q)]^{-\gamma(p,q)}. \label{distribution}
\end{equation}

\begin{figure}
\includegraphics[width=14cm]{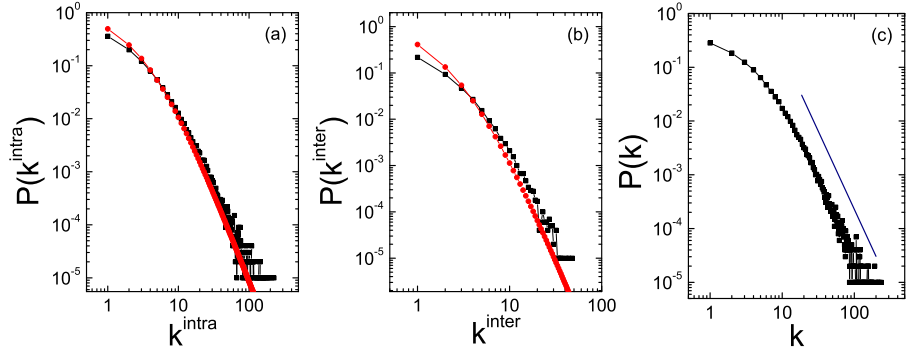}

\caption{(Color online) Log-log representation of distributions of
intra-community degree (a), inter-community degree (b), and total
degree (c) of nodes. All the simulation results (squares) display
good right-skewed distributions. The circles in (a) and (b) denote
analytical results predicted by Eqs. (\ref{degreeintra}) and
(\ref{degreeinter}), respectively. The solid line in (c) is guide to
the eye with power-law decay exponent $\gamma=3.0$. The experiment
network has a total number of nodes $N=10^{5}$ with parameters
$M=10$, $n=5$, $p=0.4$, and $q=0.4$, respectively.} \label{fig1}
\end{figure}

\begin{figure}
\includegraphics[width=10cm]{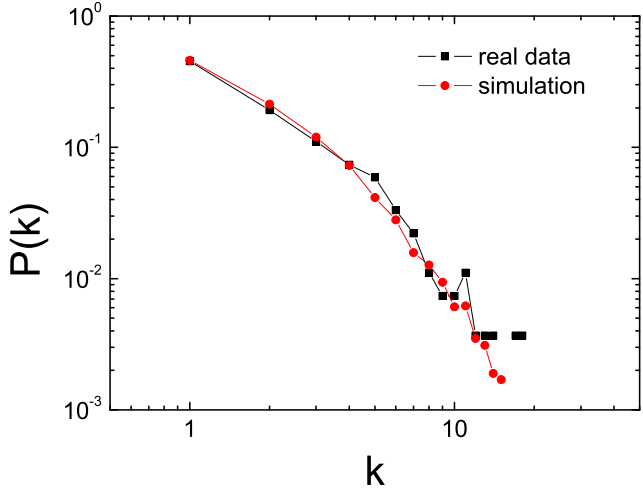}

\caption{(Color online) The degree distribution of econophysicists
(squares) of an econophysics scientific collaboration network
\cite{Zhang_06}. The circles correspond to computer simulations of
our model with parameters $M=10$, $n=2$, $p=0.4$, and $q=0.4$,
respectively.} \label{fig2}
\end{figure}

In Fig. \ref{fig1} we present numerical results of distributions of
the intra-community degree, the inter-community degree, and the
total degree of nodes in log-log scale. The experimental network is
generated by the proposed scheme with $N=10^{5}$, $M=10$, $n=5$,
$p=0.4$, and $q=0.4$, respectively. The distributions of the
intra-community degree and the inter-community degree, shown in
Figs. \ref{fig1}(a) and \ref{fig1}(b), agree with analytical results
of Eqs. (\ref{degreeintra}) and (\ref{degreeinter}), respectively.
The small deviations between computer simulations and analytical
solutions at both ends of the distributions appears to be the
mathematical approximation of the boundary conditions and the finite
size effect due to the relatively small network sizes used in the
simulations. According to the evolving rule of our network, nodes
with larger intra- (or inter-) degree have higher probability to
gain new links, then the usual degree preferential attachment is
reasonably kept. This means that the right-skewed character of the
network, such as the node's total degree, will retain. As shown in
Fig. \ref{fig1}(c), the total degree distribution of nodes is well
expected showing a good right-skewed character, which is reasonably
in agreement with the condition of many realistic systems
\cite{Amaral_97}.

To illustrate the predictive power, we also compare the numerical
result of our network with the statistics of an econophysics
collaboration network. In the econophysics collaboration network,
each node represents one scientist. If two scientists have
collaborated one or more papers, they would be connected by an edge.
Zhang \emph{et al.} took the largest connected component of this
network, which includes $271$ nodes and $371$ edges, and provided
the best division, i.e., $M=10$ \cite{Zhang_06}. In Fig. \ref{fig2}
we plot the degree distribution of econophysicists of the
econophysics collaboration network which is fitted by computer
simulations of our network starting with $10$ communities. To gain
$p$ and $q$, we fit the connectivity distribution $P(k)$ obtained
from this collaboration network with Eq. (\ref{distribution}),
obtaining a good overlap for $p = 0.75$ and $q=0.15$ (Fig.
\ref{fig2}).

\section{Conclusion}

Networks with community structures underlie many natural and
artificial systems. It is becoming essential to model and study this
kind topological feature. We presented a simplified mechanism for
networks organized in communities, which corresponds to local events
during the system's growth. The generated network is highly
clustered and has a good rightskewed distribution of connectivity,
which have been found very common in most realistic systems. The
present paper only suggests a simple way for generating community
networks. The shape of the resulting network is deterministic in
some extent. It is more interesting to model the evolution of
communities, especially the self organization (or emergence) of
communities in the natural world \cite{Kumpula_07}, e.g., expansion
and shrinkage, which is left to future work.

\section{Acknowledgements}

The authors acknowledge financial support from NSFC/10805033,
SocialNets/217141, STCSM/08ZR1408000, PTDC/FIS/71551/2006, and
FCT/SFRH/BPD/30425/2006.

\end{document}